\documentclass[fleqn,usenatbib]{mnras}

\usepackage{mathptmx}

\usepackage[T1]{fontenc}

\DeclareRobustCommand{\VAN}[3]{#2}
\let\VANthebibliography\thebibliography
\def\thebibliography{\DeclareRobustCommand{\VAN}[3]{##3}\VANthebibliography}
\defcitealias{Whitworth2023}{Paper~I}
\defcitealias{Vazquez2024}{VGG24}

\usepackage{graphicx}	
\usepackage{amsmath}	
\usepackage{lmodern}
\usepackage{caption}
\usepackage{subcaption}
\usepackage{amssymb}
\usepackage{array} 

\usepackage{ulem}

\usepackage{bm}
\usepackage[dvipsnames]{xcolor}

\newcommand{\epsff} {\epsilon_{\rm ff}}
\newcommand{\Msun} {M_\odot}
\newcommand{\pcc} {{\rm cm}^{-3}}
\newcommand{\sfrmax} {SFR$_{\rm max}$}
\newcommand{\tff} {\tau_{\rm ff}}

\definecolor{myblue}{RGB}{0, 90, 189}
\definecolor{mygreen}{RGB}{0, 80, 0}

\newcommand{\mgt} {\color{magenta} }
\newcommand{\mst} {\color{magenta} \sout}

\title[How do SNe regulate the SFR?]{How do supernovae 
regulate the star formation rate in dwarf galaxies?}

\author[D. J. Whitworth et al.]{
D. J. Whitworth,$^{1,2}$\thanks{E-mail: david.whitworth@ens-lyon.fr}
E. V\'azquez-Semadeni,$^{1}$\thanks{E-mail: e.vazquez@irya.unam.mx}
J. Ballesteros-Paredes, $^{1}$
G. C. G\'omez, $^{1}$
\\
$^{1}$Universidad Nacional Autónoma de México. Instituto de Radioastronom\'{i}a y Astrof\'{i}sica. Apdo. Postal 72-3 (Xangari)\\, Morelia, Michoc\'{a}n 58089, M\'{e}xico\\
$^{2}$Centre de Recherche Astrophysique de Lyon UMR5574, ENS de Lyon, Univ. Lyon1, CNRS, Université de Lyon, 69007, Lyon, France\\
}
\date{Accepted XXX. Received YYY; in original form ZZZ}

\pubyear{\the\year{}}

\begin{document}
\label{firstpage}
\pagerange{\pageref{firstpage}--\pageref{lastpage}}
\maketitle

\begin{abstract}
Stars form in cold, dense clouds embedded in galactic discs, but whether their formation is primarily regulated by gravitational collapse, turbulence, or stellar feedback remains unclear. Using four high-resolution dwarf galaxy simulations with and without supernova (SN) feedback and magnetic fields, we test how feedback regulates the supply of dense gas and, consequently, the star formation rate (SFR). Although the SFR does increase when SNe are turned off, this increase is only by a factor of a few. Instead, across all models, the theoretical maximum SFR originally proposed by Zuckerman and Palmer, defined as the ratio of the total dense gas mass to its mean free-fall time (${M_{\rm dense}}/{\tff}$), always exceeds the measured SFR by nearly two orders of magnitude. Moreover, the increase of the SFR in the case without SNe is accompanied by a nearly corresponding increase of the total dense gas mass ($M_{\rm dense}$), such that the dense-gas depletion time, $\tau \equiv {\rm SFR}/M_{\rm dense}$, decreases by only $\sim 33\%$ in the hydrodynamical case and by about 55\% in the magnetohydrodynamical models. This indicates that SN feedback does not primarily act by slowing the collapse of dense gas, but instead by limiting how much diffuse gas can be converted into dense gas. 
\end{abstract}




\section{Introduction}

One of the fundamental problems concerning star formation (SF) in galaxies was posed decades ago by \citet{Zuckerman1974}, who pointed out that, if all molecular clouds in the Milky Way were undergoing free-fall collapse, the resulting Galactic star formation rate (SFR) would be much larger than the observed value. They estimated this theoretical maximum SFR (hereafter labelled \sfrmax) as
\begin{equation}
    \hbox{\sfrmax} \approx \frac{M_{\rm mol}}{\tff(\bar\rho)},
    \label{eq:sfrmax}
\end{equation}
where $M_{\rm mol}$ is the total mass in molecular gas in the Galaxy, and 
\begin{equation}
    \tff(\bar\rho) \equiv \left(\frac{3 \pi}{32 G \bar\rho} \right)^{1/2}
    \label{eq:tff}
\end{equation}
is the free-fall time for the mean density $\bar\rho$ of the molecular gas. For a total molecular gas mass $M_{\rm mol} \approx 10^9 \Msun$ \citep{Heyer_Dame15} and a mean number density $n \approx 100\, \pcc$, one finds $\hbox{\sfrmax} \sim 300\, \Msun\, \hbox{yr}^{-1}$, roughly two orders of magnitude larger than current observational estimates of the SFR of $\sim 2\, \Msun\, \hbox{yr}^{-1}$ \citep{Chomiuk_Povich11}. 

Such a large discrepancy between the theoretical maximum SFR and the observed value, to which we refer as {\it the star formation conundrum}, has been in general attributed to the effect of stellar feedback on clouds, either by driving turbulence that prevents their collapse and maintains them in approximate virial equilibrium or causes them to be gravitationally unbound \citep[e.g.,] [] {Zuck_Evans74, MacLow_Klessen04, McKee_Ostr07, Padoan2016, Evans2021, Grudic2022, Kim2022, Ballesteros2024, Krumholz2025}, or by destroying collapsing molecular clouds long before most of their mass has been converted into stars \citep[e.g.,] [] {Field70, Franco+94, Colin+13, MacLow+17, Vazquez2019}.

However, the precise mechanism by which stellar feedback can reduce the SFR is not fully understood. Supernova (SN) feedback on molecular clouds is often considered the main energy-injection mechanism to interstellar turbulence from massive stars \citep[e.g.,] [] {MacLow_Klessen04, Padoan2016}, and indeed, it has been found to significantly reduce the SFR in comparison to simulations not including it \citep{Khullar2024}. However, how this compares to the \sfrmax\ estimate was not considered by those authors. Moreover, the effect of SN feedback on clouds appears to depend strongly on the distance between the SN and the cloud \citep{Iffrig2015}, and the net driving efficiency of ISM turbulence by SNe appears to be rather low, between 1.5 and 4.5\% \citep{Iffrig2017}. Furthermore, \citet{Brucy+20, Brucy+23} have suggested, through kiloparsec-scale numerical simulations, that in high-surface density galaxies, stellar feedback is insufficient to reduce the SFR to realistic levels, and that additional sources of turbulence, forms of support, or effects may be necessary in these cases. 

In addition, pre-SN photoionizing radiation from massive stars is also an important agent in regulating the SFR, since they act {\it inside} the clouds, in contrast to the effect of SNe, which typically explode {\it outside} the clouds \citep{Geen2016,Vazquez2019,Chevance2022,Bending2022}. Thus, while SN explosions can contribute to either cloud formation by external compression \citep[e.g.,] [] {Inutsuka2015} or to cloud erosion by stirring their outskirts \citep{Iffrig2015}, photoionizing feedback works exclusively to destroy the clouds \citep[e.g.,] [] {Matzner2002, Haid2019}. In conclusion, the precise effect of stellar feedback is not well constrained yet.

Magnetic support is one of the most frequently invoked mechanisms (other than feedback) to reduce the SFR \citep[e.g.,] [] {Hennebelle2019, Krumholz2019}, but, on the other hand, \citet{Zamora2018} have shown that moderate-intensity fields can actually {\it speed up} star formation by inhibiting turbulence generation during cloud assembly, and thus allowing faster onset of collapse. Additionally, \citet{Whitworth2023} have recently found that magnetic fields do not seem to reduce the global SFR in low-metallicity dwarf galaxies. Therefore, the precise effect of magnetic fields in the regulation of the {\it global} SFR also remains unclear.

Of course, a crucial form of support for the gas on scales larger than a few hundred parsecs in spiral galaxies is the rotation itself, although it is well known that it has to be complemented by significant on-plane velocity fluctuations in order to be fully stabilized \citep[e.g.,] [] {Safronov60, Toomre64, Rafikov2001}. In this paper we contribute to the understanding of the role of various mechanisms by comparing the relative contribution of rotation, feedback-induced turbulence, and magnetic fields to the regulation of the SFR in the disks of dwarf galaxies, using numerical simulations where the different processes are alternatively turned on and off. The plan of the paper is as follows: Section \ref{methods} discusses the set up of the simulations used in this study. Section \ref{results} presents the results of our analysis. Section \ref{discussion} analyses our results in the context of the dense gas reservoirs and star formation and in Section \ref{conclusion} we summarise our key findings.

\section{Methods}
\label{methods}

\subsection{Numerical Modelling}
\label{sec:Modelling}

For this work we use the quasi-Lagrangian moving mesh code \textsc{arepo} \citep{Springel2010}. It solves the ideal magnetohydrodynamic (MHD) equations on a 3D unstructured Voronoi mesh. This allows the Voronoi cells to trace the local gas velocity \citep{Pakmor2011}. After each time step the cells are restructured. By doing this, the code allows for a continuous and adaptive resolution whilst keeping the mass in the cells relatively constant, only varying by up to a factor of two. We use a \citet{Jeans1902} length criterion to ensure this length is always resolved with at least 8 cells \citep{Truelove1997,Federrath2011}. This ensures we avoid artificial fragmentation and makes sure the collapsing gas is fully resolved. We employ the \cite{Powell1999} divergence cleaning method along with an HLLD Riemann solver \citep{Miyoshi2005} in order to mitigate the divergence errors that arise in MHD models. For a more detailed discussion on \textsc{arepo} we point to \citet{Springel2010}, \citet{Pakmor2011}, and \citet{Whitworth2023}, which described the specific models used in this work in more detail. However, below we present a brief discussion on the setup used.

The ISM model implemented here is the same as in \citet{Whitworth2022} and \citet[] [hereafter \citetalias{Whitworth2023}] {Whitworth2023}. It uses a non-equilibrium, time-dependent, chemical network based on the work of \cite{Gong2017}, which includes cosmic-ray ionization. The self-shielding of dissociating ultraviolet (UV) radiation is modelled using the TreeCol algorithm of \cite{Clark2011} with a shielding radius of 30\,pc. We take the molecular and atomic cooling functions from \cite{Clark2019}, which were originally described by \cite{Glover2010}. We are able to directly trace nine non-equilibrium species (H$_2$, H$^+$, C$^+$, CH$_x$, OH$_x$, CO, HCO$^+$, He$^+$ and Si$^+$) and through conservation laws derive a further eight (free electrons, H, H$_3^+$, C, O, O$^+$ He and Si).

We model star formation with an accreting sink particle model where each sink represents a star forming region of gas. For a sink to form, a number of formation criteria must all be met to ensure that the gas is 
collapsing and self gravitating. 
A sink is only formed if gas exceeds 
a density threshold of $\rho_{\rm c}$, of $2.4 \times 10^{-21} \: {\rm g  \: cm^{-3}}$ (equivalent to $n = 10^3{\rm \: cm^{-3}})$ and within an accretion radius, $r_{\rm acc}$, of 1.75\,pc and also satisfies the following criteria:

\begin{enumerate}
  \item The gas flow must be converging with negative divergence of both velocity and acceleration.
  \item The region must be centred on a local potential minimum.
  \item The region must not fall within the accretion radius of another sink particle.
  \item The region must not move within the accretion radius of other sink particles in a time less that the local free-fall time.
  \item The region must be gravitationally bound.
\end{enumerate}

If all criteria are met, then the cell(s) form a sink particle of mass up to $200\, \rm M_\odot$, with accreted mass from a cell limited to $90\%$ of that cells mass.

Once formed, sinks 
continue to accrete gas from the surrounding cells only if a cell is within $r_{\rm acc}$, exceeds 
$\rho_{\rm c}$ and is gravitational bound (i.e. 
in falling due to self-gravity). 
We discretise the mass in the sink to determine a stellar populations based on the method presented in \cite{Sormani2017}, assuming a 
star formation efficiency within the sink of $\epsilon_{\rm SF} = 0.1$ . The remainder is treated as unresolved gas. This subgrid approach accounts for the limited resolution of the individual stars and known inefficiency of star formation at sink-creation densities \citep{Krumholz2005,Evans2009}. 

If there are 
stars of mass greater than $8M_\odot$ within a sink, once they reach the end of their life they generate a supernova explosion (SNe). No other form of feedback is considered in this paper. The SN injection routines are described in detail in \cite{Tress2019} and \cite{Whitworth2022} but we briefly describe them here. Taking the value of $\epsilon_{\rm SF}$ and applying it to the sink mass, we determine the total stellar mass within the sink. From this a stellar population is assumed using the Salpeter IMF \citep{Salpeter1955} and a set of discrete mass bins is populated via a Poisson distribution. 
The energy from a SN is injected over a radius $R_{\rm inj}$, 
set to 16 cells. If $R_{\rm inj}$ is greater than the radius of the remnant at the end of its Sedov-Taylor phase ($R_{\rm ST}$), i.e. if $R_{\rm ST} < R_{\rm inj}$, we inject energy into the ISM by setting the temperature in the bubble to 10$^4$\,K, and let the ionisation state of the gas to evolve from the injected temperature. If, however, $R_{\rm ST} > R_{\rm inj}$, then we directly inject $10^{51}$~erg into the gas as thermal energy and fully ionise the gas. We only include type IIa supernova and therefore are likely to slightly under predict the rates compared to the expected rates in Magellanic clouds of 2.5-4.6 SNe per millennium \citep{Maoz2010} and an expected one type Ia SN for every 4-6 type IIa core collapse SNe.

We 
inject mass evenly distributed into the surrounding gas at the scale of the energy injection via $\rm M_{\rm in} = (\rm M_{\rm sink} - M_{\rm stars})/nSN$, where M$_{\rm in}$ is the injected mass, M$_{\rm sink}$ is the mass of the sink at the time of the SNe, M$_{\rm stars}$ is the mass of stars in the sink and nSN is the number of remaining SNe that the sink will produce. After the last SN has occurred, the sink is considered to be fully-stellar mass and is converted into a collisionless $N$-body star particle. If a sink is unable to produce SNe due to being populated with low mass stars, then after 10\,Myr, a fraction, $\epsilon_{\rm SF} = 0.1$, of its mass is turned into a star particle with the rest being returned as gas to the surrounding cells. This mass return is exclusively tied to the feedback routine. If the routine is disabled, as in the two no-SNe models, the sinks continue to accrete mass indefinitely and no mass is returned back to the ISM. We do not include the effects of winds or jets.

For a more detailed description of the code and our custom physics modules, see \citetalias{Whitworth2023} and \cite{Whitworth2022}.

\subsection{Initial Conditions}
\label{sec:ICs}

For the models presented in this work we take our initial conditions as the 930\,Myr snapshot from model MHD\_SAT presented in \citetalias{Whitworth2023}. Here we describe the set up for that model and in the following section the changes made for the 3 new simulations completed for this work.

The initial, isolated, low metallicity dwarf galaxy is modelled as a disc consisting of two components: a dark matter halo and a gaseous disc. Sink particles are formed over time with no initial sink or star particle distribution implemented. By $930$\,Myr the sinks that are present have formed from the gaseous disc with a stellar background population having arisen from the sink particles over time. 

The dark matter halo is modelled with a spheroidal \cite{Hernquist1990} density profile:
\begin{equation}
        \rho_{\rm sph}(r) = \frac{M_{\rm sph}}{2\pi} \frac{a}{r(r + a)^3},
\label{halo_profile}
\end{equation}
where $r$ is the spherical radius, $a = 7.62$~kpc is the scale-length of the halo{\mgt ,} and $M_{\rm sph} = 2 \times 10^{10} M_\odot$ is the halo mass. The gaseous disc component is initially modelled as a double exponential density profile:
\begin{equation}
  \rho_{\rm disc}(R,z) = \frac{M_{\rm disc}}{2\pi h_z
      h^2_R
 }{\rm sech^2}  \left(\frac{z}{2h_z}\right) {\rm exp} \left( - \frac{R}{h_R} \right),
\label{gas_profile}
\end{equation}
where $R$ is the cylindrical radius and $z$ is height above the mid-plane, $M_{\rm disc} = 8 \times 10^7 M_\odot$ is the gas mass of the disc, and $h_z = 0.35$~kpc and $h_R = 0.82$~kpc are its scale-height and scale-length, respectively. See Table \ref{mass_parameters} for the initial component parameters.

The seed magnetic field is initiated as a uniform, mono-directional, poloidal seed field of strength $10^{-2}\mu$G. At 930\,Myr, the field has reached a steady state of mass-weighted $|B| \sim 10$\,$\mu$G and volume-weighted $|B| \sim 0.6$\,$\mu$G. The metallicity $Z$, dust-to-gas ratio $\xi$ (which has a Solar neighbourhood value of $\xi = 0.01$), cosmic ray ionisation rate of atomic hydrogen  $\zeta_{\rm H}$, far-ultraviolet (FUV) field strength $G$ (given in Habing units $G_0$) are all set to 10\% of their solar neighbourhood values and listed in Table \ref{parameters}. These values are constant throughout the simulation. For a more detailed description of the initial setup we refer the reader to the original work, \citetalias{Whitworth2023}.

\begin{table}
\centering
\begin{tabular}{c c c p{1cm}}
 \hline
  & Mass (M$_\odot$) & Scale length (kpc) & $h_z$ (kpc) \\ 
 \hline
 DM Halo & 2.00 $\times 10^{10}$ & 7.62 & - \\
 Gas disc & 8.00 $\times 10^{7}$ & 0.82 & 0.35 \\
 \hline
\end{tabular}
\caption{Parameters for the different galactic components}
\label{mass_parameters}
\end{table}

\begin{table}
\centering
\begin{tabular}{c c c c c c c c c}
 \hline
  Model & Z & $\log_{10} \xi$ &  $\zeta_{\rm H}$ & $G$ & $\log_{10} B$\\
            &  ($Z_\odot$) &   & (s$^{-1}$) &  ($G_0$) &  ($\mu$G)  \\
 \hline
 MHD\_SAT & 0.10 & -3 & 3 $\times 10^{-18}$ & 0.17 & -2 \\
 
\hline
\end{tabular}
\caption{Values used in the initial conditions for MHD\_SAT for metallicity $Z$, dust-to-gas ratio $\xi$, relative to the value in solar metallicity gas, cosmic ionisation rate $\zeta_{\rm H}$, UV field strength $G$, and magnetic field strength $B$.}
\label{parameters}
\end{table}

\subsection{New models}

We run three new models and continue run of MHD\_SAT for a further 75\,Myr. The new models begin from the 930\,Myr snapshot of MHD\_SAT as described above. This work aims to look at the importance of feedback and magnetic fields in the role of galactic star formation and how they affect the star formation rate (SFR). With respect to that the new models are designed to look into this. We take the snapshot at 930\,Myr and modify it to represent different regimes. First, in model \textbf{MHD\_no\_SNe} we turn off the SNe routine but allow sinks to still form and accrete. For model \textbf{Hydro} we turn off the magnetic field but keep the SNe routine turned on. In Model \textbf{Hydro\_no\_SNe} we turn off both the SNe and the magnetic field. See Table \ref{new_models}.

\begin{table}
\centering
\begin{tabular}{c c c}
 \hline
  Model & SNe & $B$\\
 \hline
 Hydro & $\checkmark$ & $\times$ \\
 Hydro no SNe & $\times$ & $\times$ \\
 MHD & $\checkmark$ & $\checkmark$ \\
 MHD no SNe & $\times$ & $\checkmark$ \\
\hline
\end{tabular}
\caption{The four models used for this work showing which ones have the supernova{\mst{e}} routine (SN) and magnetic field ($B$) turned on}
\label{new_models}
\end{table}

We run models MHD and Hydro for 150\,Myr from the snapshot at 930\,Myr. This ensures that the Hydro model has reached a new steady state in star formation. When the magnetic field is turned off, pressure support in the ISM suddenly changes and decreases, creating a short burst in star formation. These new stars then generate more SNe, causing a short dynamical change in the disc. After a few Myr the SFR again reaches a steady state. We define the steady state periods as 1000--1070 Myr for the Hydro and MHD simulations with SNe, 950--980 Myr for the Hydro\_no\_SNe simulation, and 950--995 Myr for MHD\_no\_SNe.

The models without SNe are run for as long as is computationally possible. With no feedback in the models the gas has nothing to support it from undergoing self collapse due to gravity other than the rotation of the disc and magnetic field. This collapse into dense structures causes \textsc{arepo} to hyper refine in the densest gas beyond the sink creation density and form millions of new cells. Whilst these cells only exist for a short time before being refined (see Section \ref{sec:Res}), this eventually becomes untenable in the code and it can no longer proceed. This happens after 50\,Myr in Hydro\_no\_SNe model, and after 70\,Myr in the MHD\_no\_SNe model, showing that the magnetic field does slow down the collapse of the gas into denser structures by roughly 40\%.

\subsection{Resolution}
\label{sec:Res}

We keep the resolution the same as in the original work. The base target mass resolution for the gas cells is set to $50$~M$_\odot$ using the default \textsc{arepo} refinement scheme. The scheme keeps the cell mass within a factor of 2 of this target mass as cells merge and split during refinement. We also use a Jeans refinement criterion where by we resolve the Jeans Length by at least 8 cells. This implies that, at number densities of $n = 100 \, \rm cm^{-3}$, we reach an average cell radius ($r_{\rm cell}$) of $~ 0.79$~pc and, at $n = 10^3$\,cm$^{-3}$, $r_{\rm cell}$ is $~0.31$~pc. Due to the fact that the cells range in both mass and size we adopt an adaptive softening for the gas cell size. In the initial conditions for the parent simulation we applied a fixed softening length to both the dark matter (64\,pc) and the sink particles (2\,pc).

\begin{figure}
	\includegraphics[width=\columnwidth]{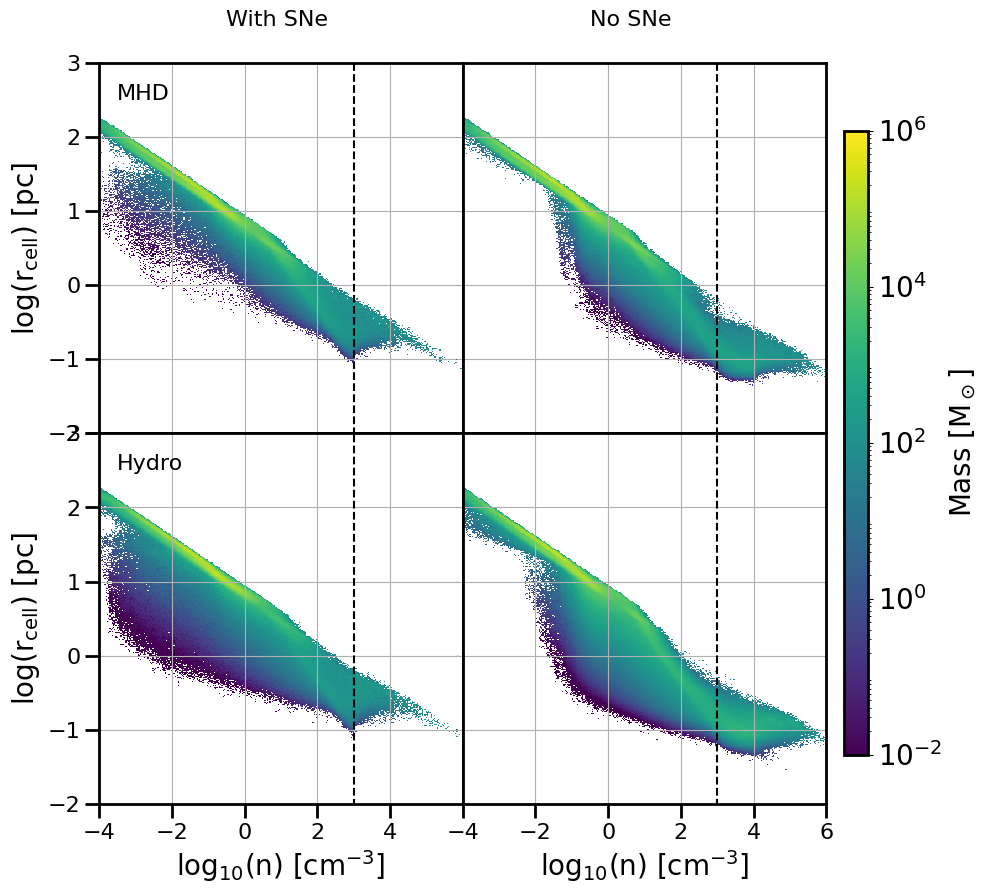}
    \caption{The mass weighted cell radius against number density for each of the simulations taken at the last snapshot. The dashed vertical line represents the sink creation density threshold. We can see here the increase in hyper refined cells in the models without supernova feedback.}
    \label{resolution}
\end{figure}

One issue that arises with the sink-creation scheme used in our simulations is that it allows for cells to go beyond the sink creation density ($10^3 \rm cm^{-3}$, the dashed line in Figure \ref{resolution}) depending on when the sinks form in the time-stepping process. These cells will typically persist for a few 
timesteps, before they are refined or accreted onto the sink. This is due to the the fact that the sink creation density is considered when the checks for creation begin and not a hard limit on them forming at this density. We can see this effect in Figure \ref{resolution}, which shows the existence of a significant number of cells with densities larger than the sink-formation threshold. 
When SNe are removed there is little to disrupt the gravitational collapse of the dense gas resulting in 
an $\sim 85\%$ increase in the number of cells above the creation density. This results in the models becoming computationally very expensive. Note, however, that this only implies that the simulations need to be stopped due to very small timesteps, but not that the collapse flow is somehow delayed in the physical time of the simulation. 

Our steady-state analysis uses periods (1000--1070 Myr Hydro and MHD; 950--980 Myr Hydro\_no\_SNe; 950--995 Myr MHD\_no\_SNe) during which the SFR, $M_{\rm gas}$, and depletion times show no large-scale variation over small time bins (See Section \ref{sec:sfr}).

\section{Results}
\label{results}

We present our results for each model in the following sections. We first look at the morphology of the gaseous disc and differences between them. We then present the magnetic field results and how they change when feedback is removed. Then, we focus on the SFR and compare it to the free-fall star formation rate, Equation \eqref{eq:sfrmax}, the theoretical maximum SFR for a dense gas cloud. 
All results, unless otherwise stated, are taken from the final snapshots of each model, Hydro and MHD at 1070\,Myr, Hydro\_no\_SNe at 980\,Myr and MHD\_no\_SNe at 998\,Myr.

\subsection{Morphology}

\begin{figure*}
	\includegraphics[width=\linewidth]{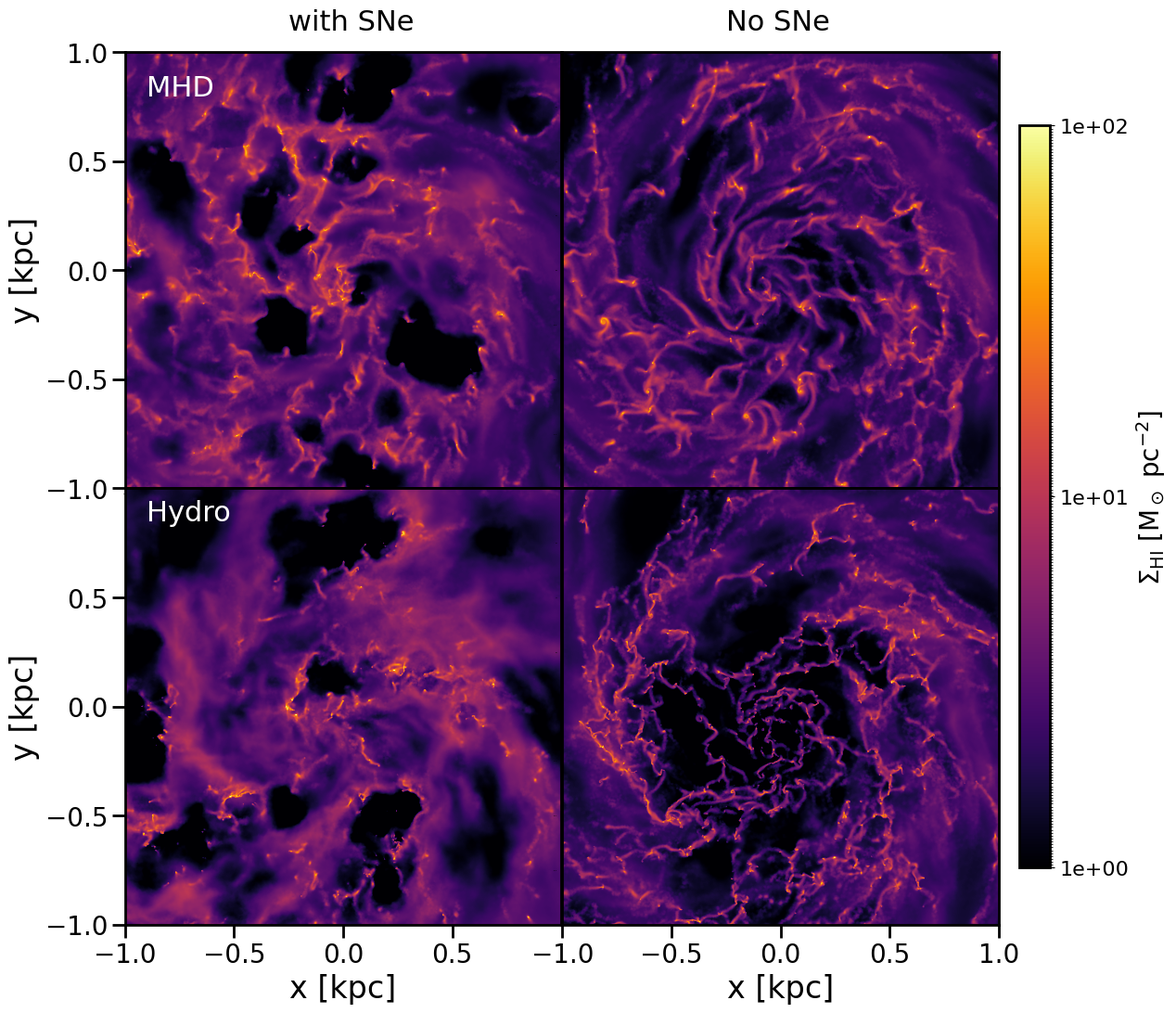}
    \caption{The HI gas surface density projections for each of the four models used in the analysis. The top row shows the hydrodynamical models, the bottom row shows the MHD models. On the left we show the models with SNe feedback, the right column has no SNe feedback. The dense filament structures that form due to self-gravity when there is no support from feedback in the gas can be clearly seen.}
    \label{fig:SD}
\end{figure*}

We plot the HI surface density projections in Figure \ref{fig:SD} for each model. The models with SNe, left hand plots, show a 
less filamentary structure, with more extended diffuse-gas regions across the projection compared to models without SNe, especially in the Hydro model. However, the MHD model does show some dense filamentary-like structures, especially in the upper left quadrant. Whereas the Hydro model shows much fewer filamentary structures with less variation in the density when SNe are included. The models that exclude SNe feedback, the right hand plots, show more filamentary and spiral-like structures and voids. 
We also see that the MHD model has fewer large-scale voids compared to the Hydro model. 

\begin{figure*}
	\includegraphics[width=\linewidth]{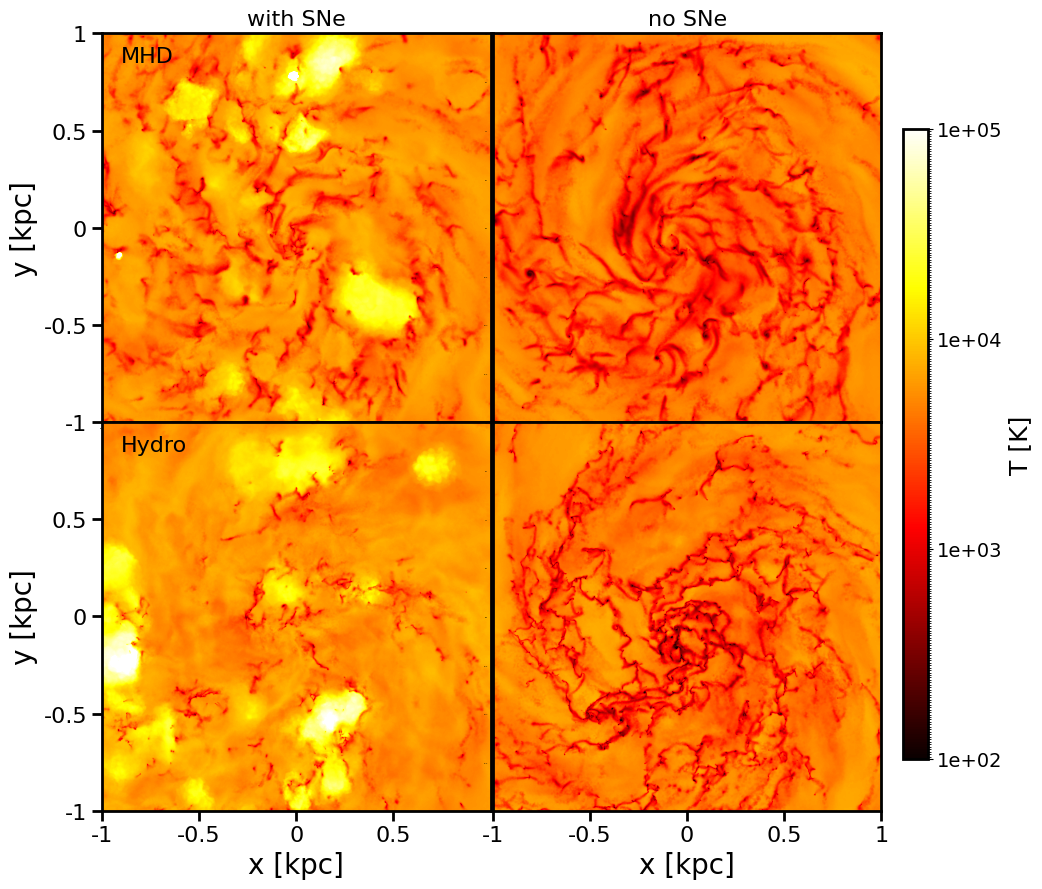}
    \caption{Integrated temperature projection of the 4 models. The diffuse voids seen in the SNe models in Figure \ref{fig:SD} are seen as hot bright spots. Whereas the voids in the models without SNe are more consistent with the diffuse gas.}
    \label{fig:temp_projection}
\end{figure*}

If we look at Figure \ref{fig:temp_projection} we can see that the voids are hot in the models with SNe, whilst when SNe are removed this hot gas obviously disappears. However, we can see that the cold gas in the Hydro model without SNe is much more sharply defined, whilst in the MHD the darker, colder regions are less well defined. So the magnetic field does not appear to be preventing thermal condensation but slowing the condensation of the gas into fine, dense, structures, supporting the idea that the field supports the gas against gravitational collapse. 

\begin{figure}
	\includegraphics[width=\columnwidth]{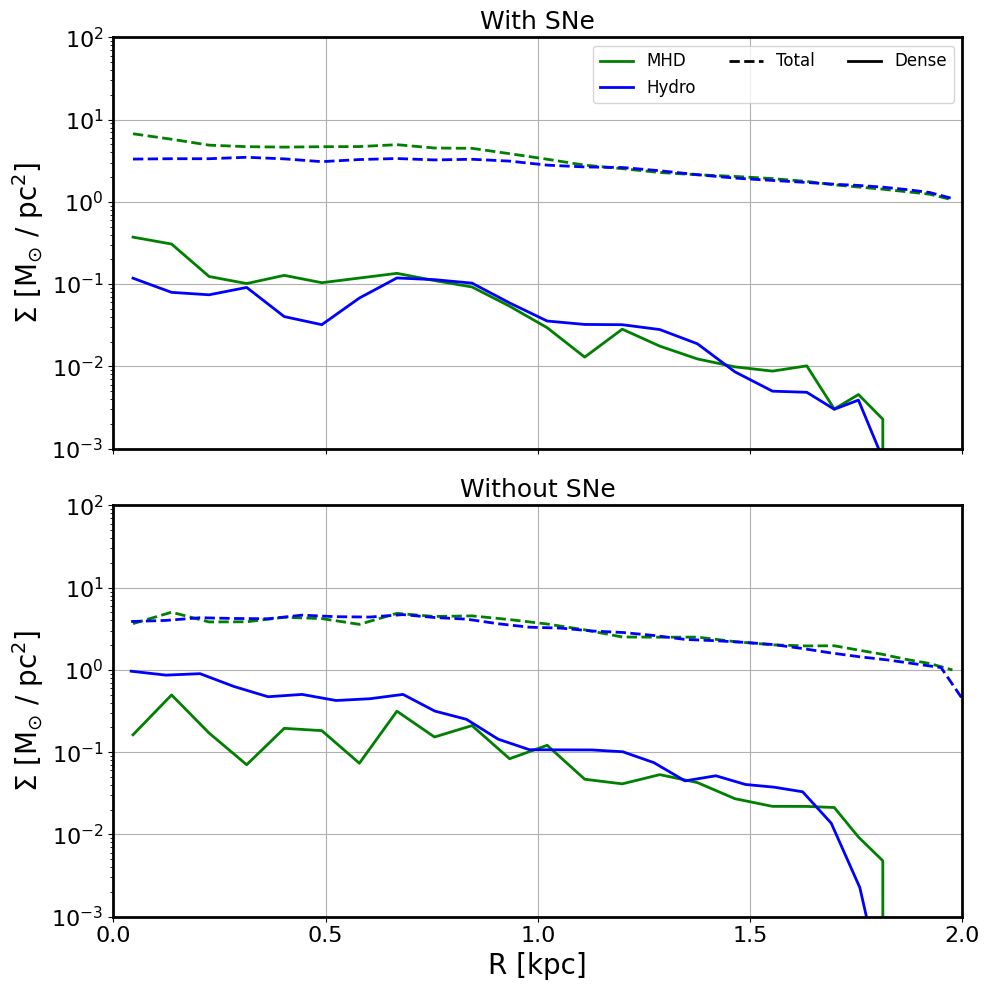}
    \caption{The total gas (dashed line) and dense gas (solid line) surface density radial profiles for the MHD (green) and Hydro (blue) simulations with and without SNe averaged over the steady-state period.}
    \label{fig:SDRP}
\end{figure}

Figure \ref{fig:SDRP} shows the time averaged total gas and dense gas, $n \geq 100$\,cm$^{-3}$, azimuthally-averaged surface density radial profile for each model. There is little change in the total gas surface density with and without SNe, except for a very small increase in MHD and small decrease in Hydro when SNe are present. When SNe feedback is removed we see that the dense gas surface profile for both models increases, especially in the Hydro case, which increase by an order magnitude. In the MHD case it is similar in the central 0.5\,kpc, but then stays higher to the point where is drops off at $\sim 1.75$\,kpc.

\subsection{Magnetic Field}
\label{sec:fields}
Figure \ref{fig:field} shows the magnetic field strength and field lines in both MHD models. The field is more organised in the model without SNe, exhibiting more stable spiral arm-like structures in the field. The field strength is also more evenly distributed across the disc with fewer low strength voids within it. This is expected as we are using ideal-MHD where the field traces the gas. As no SNe are creating large bulk motions in the gas, the field becomes more uniform since the gas distribution is less turbulent.

Figure \ref{fig:field_growth} shows the temporal evolution of the volume-weighted mean field strength in the disc of each MHD simulation, where the disc is defined as the volume within $r = 1.75$ kpc and $z = {\pm} 0.3$ kpc over time. Both models increase in field strength at the start, but the model without SNe increases at a faster rate, from $0.71\, \mu$G to $0.93\, \mu$G, whilst the simulation with SNe only reaches $0.80\, \mu$G. We cannot give a steady-state averaged volume-weighted field strength for the model without feedback, as it does not appear to be in a steady state in terms of field strength. However, the final snapshot has a volume-weighted field strength of $0.90\, \mu$G. On the other hand, the model with SNe {\it is} in a steady state, having been simulated for a much longer time \citetalias{Whitworth2023}. For consistency with the star formation rate and other analysis in this work we take the steady state from 1\,Gyr and report a volume-weighted average field strength of $0.77\, \mu$G.

Figure \ref{fig:B-n} shows the $B$-$n$ relationship for each model. {The black dotted line represents the volume-weighted average $B$ for the simulations; the solid black line shows the result from \citet{Crutcher2010}, whilst the red lines show the results from \citet{whitworth2025}. In the model with no feedback there are no SNe driving the hot, diffuse, turbulent gas, which is important in driving the growth of the small-scale dynamo \citep{Gent2021,whitworth2025}. We see that the simulations do not follow the observational predictions, especially in the diffuse gas; this is discussed in more detail in \citet{whitworth2025}, where the importance of resolution is noted. The lack of very diffuse gas in the model without feedback is a result of a lack of SNe creating hot diffuse gas.

\begin{figure*}
	\includegraphics[width=\linewidth]{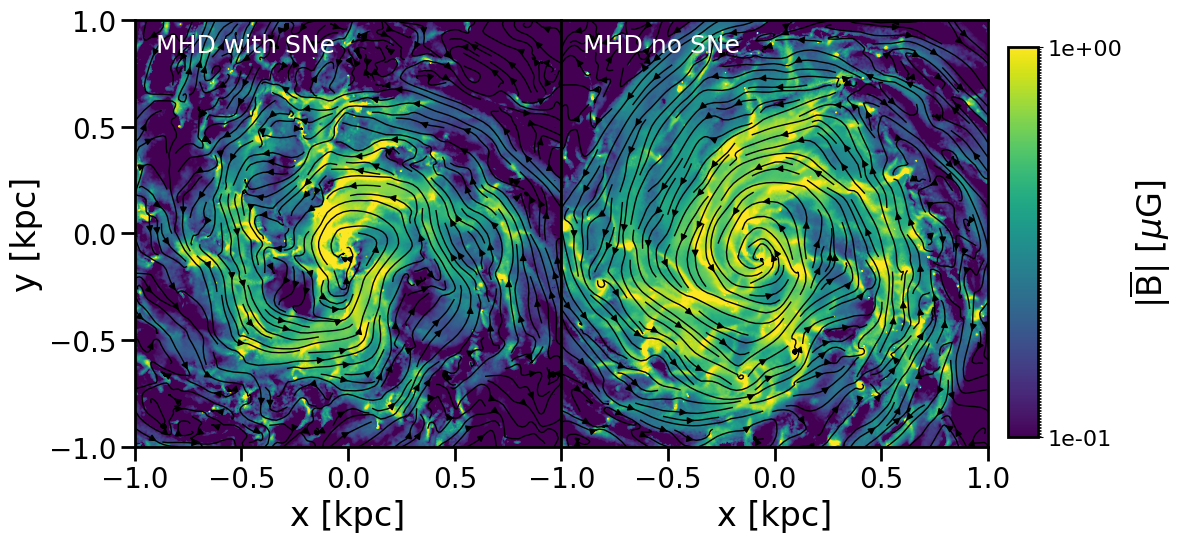}
    \caption{The strength of the magnetic field with the field line structure over plotted in the two MHD models presented in this work. The left plot is the model with SNe feedback, the right has no SNe. It is clear in these plots that without feedback the resulting field is more organised and evenly distributed across the disc.}
    \label{fig:field}
\end{figure*}

\begin{figure}
	\includegraphics[width=\columnwidth]{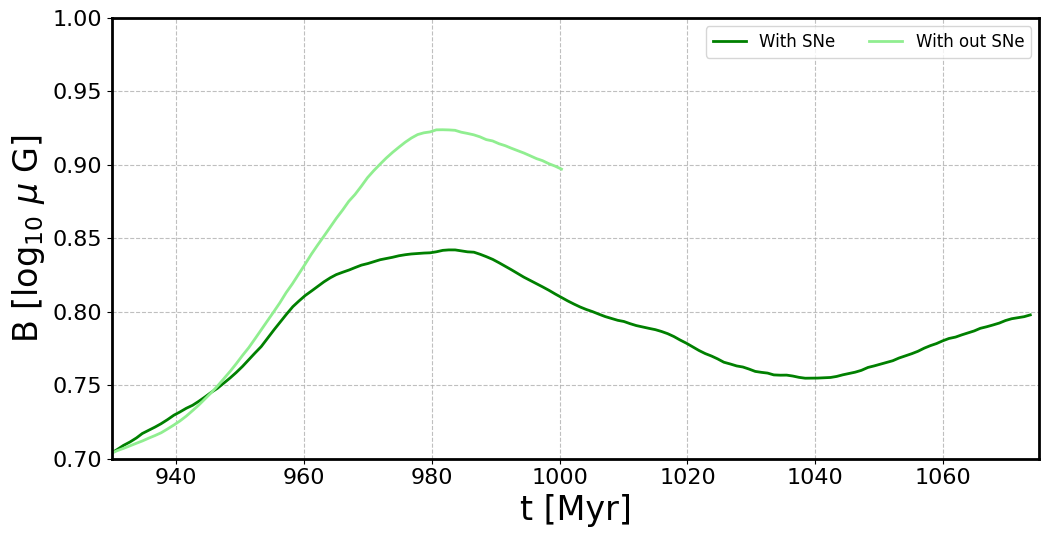}
    \caption{volume-weighted magnetic field over time of the two MHD simulations. We see the field grows when feedback is turned off but only by a small amount and appears to be decreasing at the end of the simulation.}
    \label{fig:field_growth}
\end{figure}

\begin{figure}
	\includegraphics[width=\columnwidth]{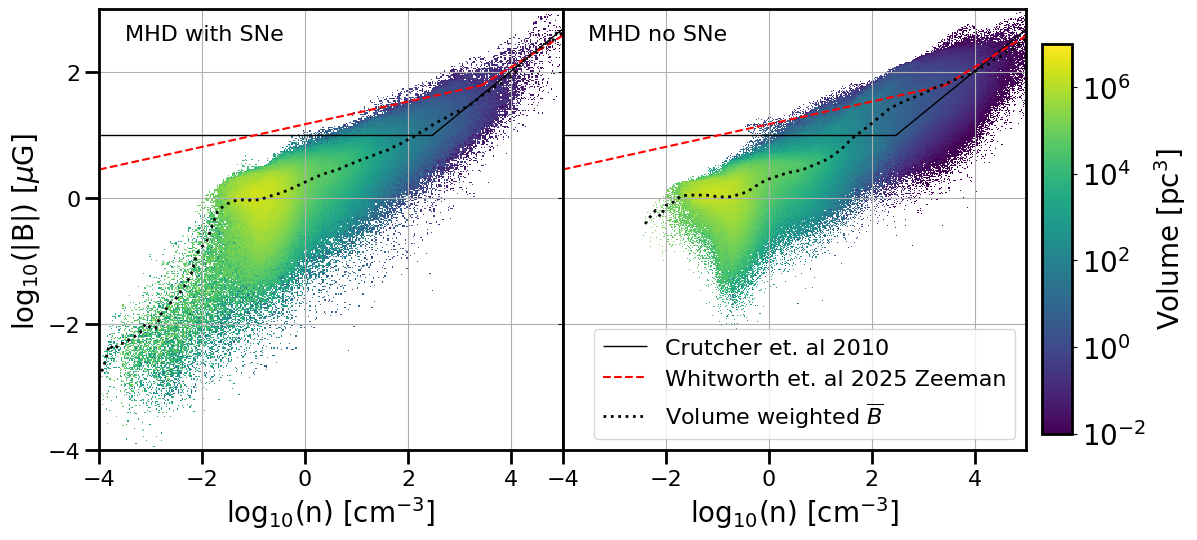}
    \caption{The $B$-$n$ relationship in the two MHD simulations. The black line represents the \protect \cite{Crutcher2010} $B$-$n$ relationship, the red dashed line is the Zeeman result from \protect \cite{whitworth2025}. The black dotted line is the volume-weighted average field strength.}
    \label{fig:B-n}
\end{figure}

\subsection{Star formation rates}
\label{sec:sfr}
To calculate the SFR we use the same method as in \cite{Whitworth2022}. 
We use post-processing to define the star formation rate between two snapshots separated by ($\delta t = t_2-t_1$) as:
\begin{eqnarray}
        {\rm SFR} = \frac {\epsilon_{\rm SF}\left(M_{\rm acc} + M_{\rm new} \right)} {\delta t}
\label{LastlyLastlyLastly}
\end{eqnarray}
where ($M_{\rm acc}$) is the gas mass accreted onto pre-existing sinks during ($\delta t$), ($M_{\rm new}$) is the mass of newly formed sinks, and ($\epsilon_{\rm SF}=0.1$) is the assumed subgrid star formation efficiency within a sink.


Following \cite{Zuckerman1974}, we define the maximum free-fall star formation rate as:}
\begin{equation}
    \hbox{\sfrmax} \approx \frac{M_{\rm gas}}{\langle t_{\rm ff} \rangle},
    \label{eq:SFRmax}
\end{equation}
where ($M_{\rm gas}$) is either the total gas mass of the disc ($M_{\rm total}$) or the dense gas mass ($M_{\rm dense}$), depending on the comparison being made. We define $M_{\rm dense}$ as:
\begin{equation}
    M_{\rm dense} = \sum_{n \geq 100 \, \rm cm^{-3}} m_i,
\end{equation}
and $\langle t_{\rm ff} \rangle$ is the mass-weighted mean free-fall time of the selected gas where
\begin{equation}
    t_{\rm ff} (\rho) \equiv \left(\frac{3 \pi}{32 G \rho}. \right)^{1/2}.
    \label{eq:tff}
\end{equation}

\begin{figure}
	\includegraphics[width=\columnwidth]{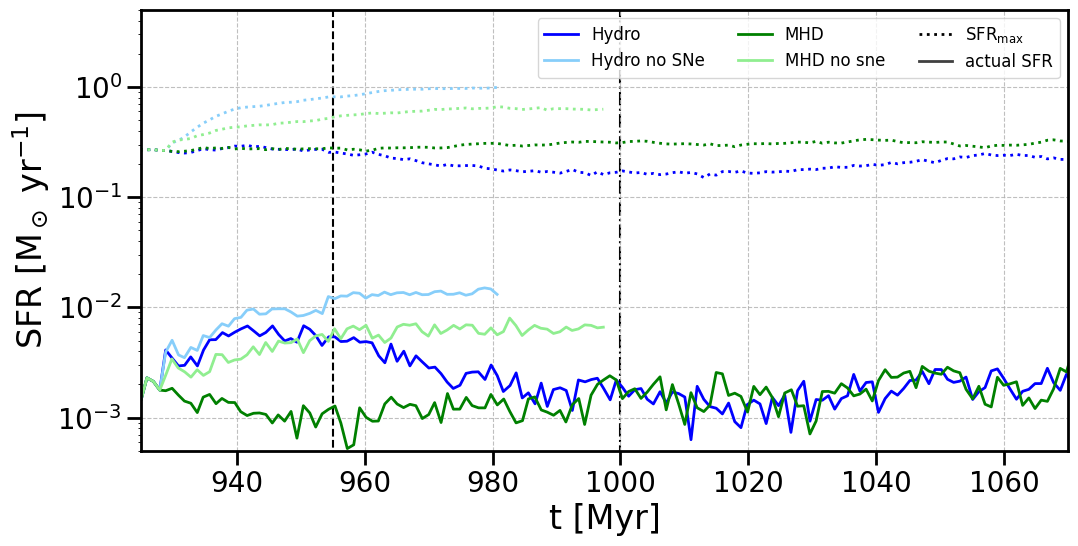}
    \caption{The star formation rate (SFR), solid lines, and the maximum free fall star formation rate (\sfrmax), dotted lines, for each of the models. We show the time range from 925\,Myr to make the divergence after the common 930\,Myr restart explicit. The black vertical dashed and dash-dotted lines indicate the starting times of the steady-state averaging periods for the no SNe and SNe runs, respectively. The averaging periods extend to the final times of each of the simulations. The starting rates for the new models are slightly higher than the initial model because the rate is based of the difference between the initial snapshot which is the same and the following snapshot where the models begin the divergence. The burst of star formation that occurs in the hydrodynamical model after we turn off the magnetic field can be seen between 930 and 950\,Myr. It then settles to a steady state at $\simeq 980$\,Myr and is similar to the MHD model. We clearly see that the \sfrmax\ is nearly 2 orders of magnitude greater than the actual SFR in all cases.}
    \label{fig:SFR}
\end{figure}

Figure \ref{fig:SFR} shows the star formation rates (SFR) for each model (solid lines) as well as the upper limit, \sfrmax \, (dotted lines). The black dashed and dashed-dotted vertical lines show the starting times of the periods we consider to be in a steady state, over which we calculate averages. These averaging periods end at the final time of each simulation. For the models without SNe this starting time is 955\,Myr, and for the models with SNe it is 1000\,Myr. We take these times as from here the models have flat SFRs. We can see a burst in star formation when the magnetic field is turned off, occurring between 930\,Myr and $\sim 980$\,Myr in the Hydro model (blue line). This also happens in the Hydro model with no feedback (cyan line), but we note that at $\sim 950$\,Myr the SFR appears to reach a steady state. Finally, we also calculate the \sfrmax, which is the theoretical maximum SFR based on the free-fall time of the gas calculated as in Equation \eqref{eq:sfrmax} where $M_{mol}$ is defined as he mass in dense gas above number densities of 100\,cm$^{-3}$, and $t_{\rm ff}$ is the mean free-fall time of the dense gas.

Table \ref{tab:SFR_table} shows the average results of the steady state periods along with the ratio between \sfrmax \, and the observed SFR, the total gas mass in dense gas (M$_{\rm dense}$) as well as the SFR$_{\rm specific}$ (the SFR per unit dense gas mass). We see that the values for \sfrmax\ are 1.7 to 1.8 orders of magnitude larger than the observed SFR in all simulations, ranging from a factor 51 in the Hydro simulation to 64 in Hydro\_no\_SNe. 

We compute the star formation efficiency per free-fall time \citep[$\epsff$;][] {Krumholz2005} for each simulation as:
\begin{equation}
        \epsff \equiv \frac{{\rm SFR}} {\left(\frac{M_{\rm gas}}{\langle\tau_{\rm ff}\rangle} \right)} \equiv \frac{{\rm SFR}} {{\rm SFR}_{\rm max}},
\label{eq:SFE}
\end{equation}
where the second identity follows from eq.\ \eqref{eq:SFRmax}, and shows that, rather than an efficiency, this quantity is actually the ratio of the measured SFR to the theoretical maximum, as discussed by \citet{Zamora+25}. We report its values for all simulations in Table \ref{tab:SFR_table}.

\begin{figure}
	\includegraphics[width=\columnwidth]{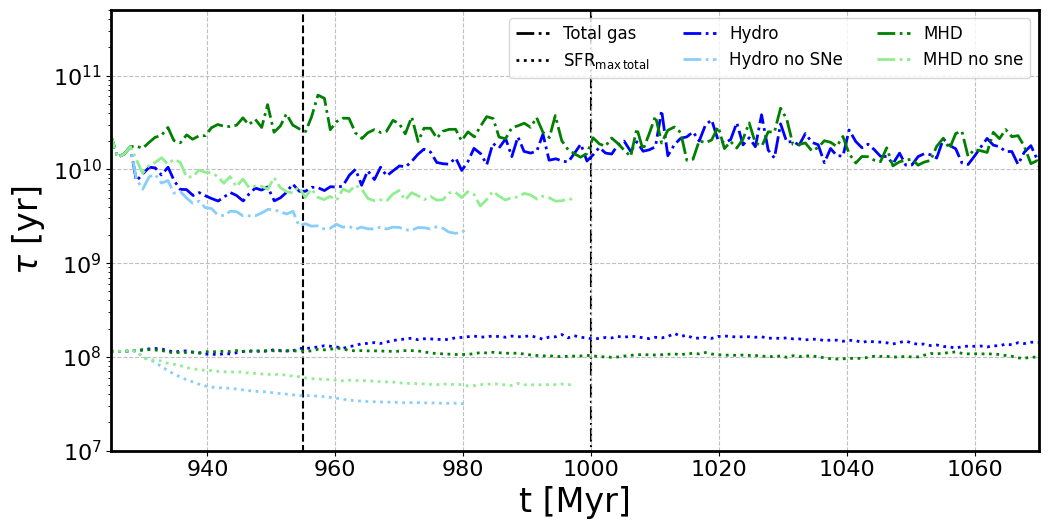}

	\includegraphics[width=\columnwidth]{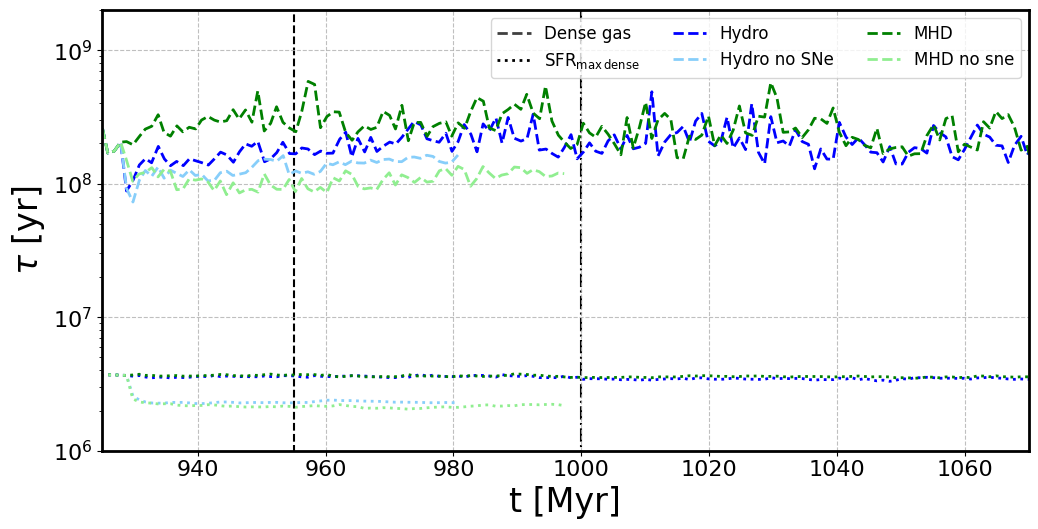}
    \caption{\textbf{Top:} Depletion times $\tau$ calculated according to eq.\ \eqref{depletion_time} using the {\it total} gas mass and either the measured SFR (dash-dotted lines) or the theoretical maximum star formation rate (\sfrmax, given by eq.\ \eqref{eq:SFRmax}; dotted lines). \textbf{Bottom:} Same as the top panel, but using the {\it dense} gas mass instead of the total. As with Figure \protect \ref{fig:SFR} we plot from 925\,Myr so that that divergence at 930\,Myr can be seen.}
    \label{fig:dep_time}
\end{figure}

Figure \ref{fig:dep_time} shows the depletion time ($\tau$) of the total gas mass (column 2) in the disc and in the dense gas (column 3). We calculate it as:
\begin{equation}
        \tau \equiv \frac {M_{\rm gas}}{\rm SFR},
\label{depletion_time}
\end{equation}
%
where  as mentioned above, $M{\rm gas}$ is either the total gas mass (column 2) or dense gas mass $M_{\rm dense}$ (column 3, $n \geq 100\,{\rm cm^{-3}}$. Note that the gas in sinks is {\it not} included in $M_{\rm dense}$. Thus $\tau_{\rm dense} = M_{\rm dense}/{\rm SFR}$ measures the timescale for the observed stellar mass production rate to deplete the dense gas reservoir, using the SFR defined above with subgrid efficiency $\epsilon_{\rm SF}=0.1$. Note that $\tau$ is the inverse of the maximum theoretical star formation rate, \sfrmax, defined in eq.\ \eqref{eq:SFRmax}. 

\begin{table*} 
\centering
\begin{tabular}{l c c c c c c c c c}
\hline
Model & SFR & 
SFR$_{\rm max}$  & $\overline{t_{\rm ff}}$ &  SFR$_{\rm max}$/SFR & $\epsff$ & $M_{\rm dense}$ & SFR$_{\rm specific}$ \\
 & (M$_\odot$yr$^{-1}$) & 
 (M$_\odot$yr$^{-1}$) & (yr) &  & & ($\geq \rm 100 cm^{-3}$) (M$_\odot$) & (yr$^{-1}$) \\ 
\hline
Hydro &          $1.74 \times 10^{-3}$ & 
0.099 & $3.45 \times 10^6$ & 56.5 & 0.018 & $3.40 \times 10^5$ & $5.09 \times 10^{-9}$ \\ 
Hydro\_no\_SNe & $12.7 \times 10^{-3}$ & 
0.792 & $2.32 \times 10^6$ & 62.1 & 0.016 & $18.3 \times 10^5$ & $7.00 \times 10^{-9}$ \\ 
MHD &            $1.86 \times 10^{-3}$ & 
0.119 & $3.58 \times 10^6$ & 63.9 & 0.016 & $4.35 \times 10^5$ & $4.35 \times 10^{-9}$ \\ 
MHD\_no\_SNe &   $6.17 \times 10^{-3}$ & 
0.315 & $2.15 \times 10^6$ & 51.0 & 0.020 & $6.74 \times 10^5$ & $9.23 \times 10^{-9}$ \\ 
\hline
\end{tabular}
\caption{Computed values for the star formation rate, the free-fall star formation rate (SFR$_{\rm max}$) based on the mean mass-weighted free fall time ($\overline{t_{\rm ff}}$) of dense gas ($n \geq \rm 100 cm^{-3}$), the mean mass-weighted free fall time of dense gas, the ratio between SFR$_{\rm max}$ and SFR, the efficiency per free-fall time ($\epsff$), the average mass of dense gas ({\it excluding} gas in sinks) and the average SFR per unit dense gas mass in each simulation. All values are taken as the time average over the defined steady-state periods.}
\label{tab:SFR_table}
\end{table*}

\begin{figure}
	\includegraphics[width=\columnwidth]{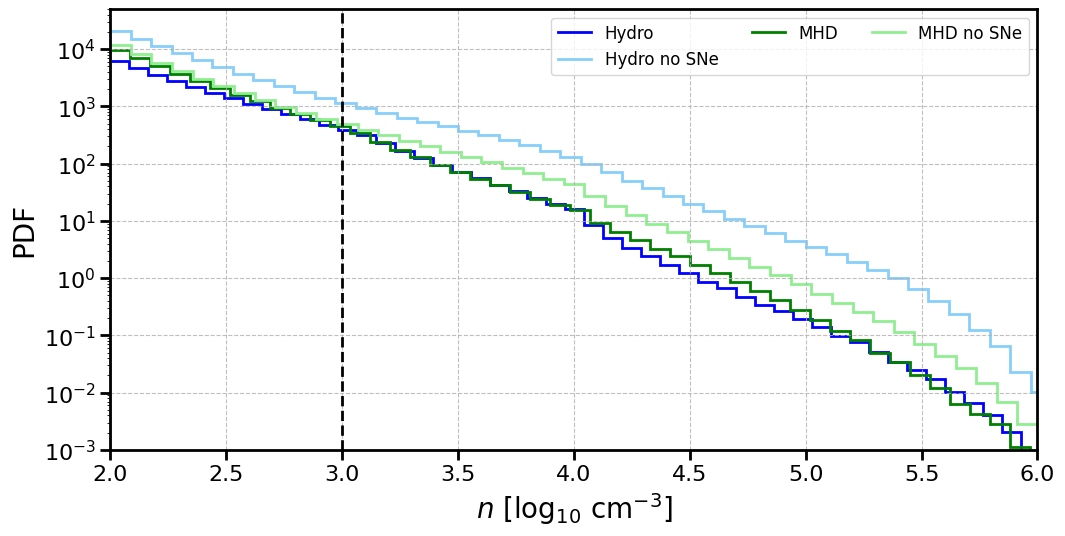}
    \caption{Time averaged, volume weighted density PDF of $M_{\rm dense}$ in each model. }
    \label{fig:den_pdf}
\end{figure}

\begin{table*}
\centering
\begin{tabular}{lccccc}
\hline
& & & Fraction of\\
      & total mass in & total mass in & total mass in & dense mass in  & dense mass in  \\
 Model & $n > 100\,{\rm cm^{-3}}$  & $100 \le n < 10^3\,{\rm cm^{-3}}$ & $n \ge 10^3\,{\rm cm^{-3}}$ & $100 \le n < 10^3\,{\rm cm^{-3}}$ & $n \ge 10^3\,{\rm cm^{-3}}$ \\
\hline
Hydro          & $4.43 \times 10^{-3}$ & $2.71 \times 10^{-3}$ & $1.71 \times 10^{-3}$ & $6.10 \times 10^{-1}$ & $3.90 \times 10^{-1}$ \\
Hydro\_no\_SNe & $19.4 \times 10^{-3}$ & $8.10 \times 10^{-3}$ & $11.3 \times 10^{-3}$ & $4.29 \times 10^{-1}$ & $5.71 \times 10^{-1}$ \\
MHD            & $4.72 \times 10^{-3}$ & $3.15 \times 10^{-3}$ & $1.57 \times 10^{-3}$ & $6.67 \times 10^{-1}$ & $3.32 \times 10^{-1}$ \\
MHD\_no\_SNe   & $7.39 \times 10^{-3}$ & $3.89 \times 10^{-3}$ & $3.50 \times 10^{-3}$ & $5.39 \times 10^{-1}$ & $4.61 \times 10^{-1}$ \\
\hline
\end{tabular}
\caption{Time-averaged mass fractions for the dense gas in each model. Columns 2--4 give the fraction of the total gas mass in gas with $n \ge 100\,{\rm cm^{-3}}$, $100 \le n < 10^3\,{\rm cm^{-3}}$, and $n \ge 10^3\,{\rm cm^{-3}}$, respectively. Columns 5--6 give the fraction of the dense gas mass ($n \ge 100\,{\rm cm^{-3}}$) in the ranges $100 \le n < 10^3\,{\rm cm^{-3}}$ and $n \ge 10^3\,{\rm cm^{-3}}$.}
\label{tab:den_mf_table}
\end{table*}

\begin{figure}
	\includegraphics[width=\columnwidth]{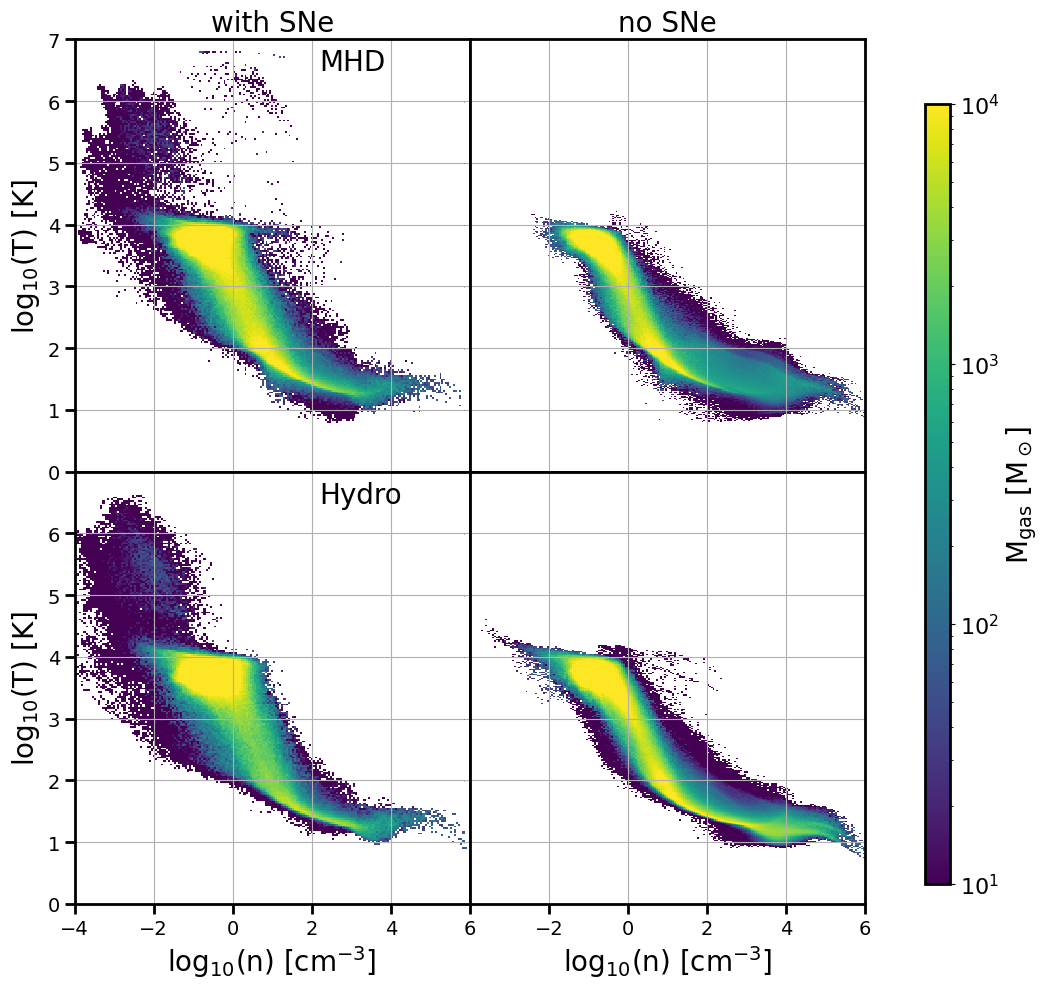}
    \caption{Number density phase plots for the gaseous discs of each simulation. We see that in the models with SN feedback the diffuse gas temperature reaches up to $10^6 \sim 10^7$\,K, where as in the models with no SN the temperature reaches maximum of $\sim 10^4$\,K with little to no gas below $n \sim 0.01$\,cm$^{-3}$. We also see a broadening of the distribution below T $\sim 100$\,K and $n \sim 1$\,cm$^{-3}$.}
    \label{fig:phase}
\end{figure}

%
%
%
%

\section{Discussion}
\label{discussion}

There are different ideas on what provides support against collapse of the dense gas and controls star formation, such as, for example, supernova driven turbulence \citep{Padoan2016} or magnetic fields providing support and reducing the SFR \citep{Hennebelle2019,Krumholz2019,Whitworth2023}. In a seminal paper, \citet{Zuckerman1974} showed that star formation is inefficient and there have been numerous works \citep[see, e.g., the review by] [and references therein] {Schinnerer2024} that have tried to help explain this. In this work we add to the discussion by looking at dense gas reservoirs and star formation in a galactic context.

\subsection{Morphology}

Focusing on the morphology of the gas in the simulations, Figure \ref{fig:SD}, we can see the physical effects that SN feedback and magnetic fields have upon the gas distribution. In both models that include feedback we see voids/bubbles and dense filamentary-like structures with a distribution of less dense material. The models with magnetic fields appear to have larger and broader filamentary structures. Once the feedback is turned off there is a distinct change in the gas distribution. We begin to see longer filamentary structures appearing to follow the rotation of the disc, forming almost spiral arm-like structures. This is especially noticeable in the Hydro simulation (bottom right plot), where we can see more numerous and larger voids (the low density dark regions) compared to the MHD model. The larger filaments 
are probably a result of the dense gas not being disrupted by the SNe or a lack of turbulent support as suggested by \citet{Padoan2016}, and the lack of heating and stirring of the diffuse gas by the SNe, so that it can freely condense to the cold phase by thermal instability. Figure \ref{fig:SDRP} shows clearly that there is more dense gas in the models without SNe, with both solid lines being consistently higher than the models with SNe. This suggests then that the role of SNe is to retard the transition from diffuse warm gas to cold dense gas by thermal condensation.

\subsection{Star formation rates and depletion times}
\label{sec:SFR_tdepl}

As would be expected, when SN feedback is disabled we see an increase in the star formation rate (Figure \ref{fig:SFR}, solid lines). Indeed, if there is nothing to prevent the dense gas from gravitationally collapsing, then the star formation rate should increase.  

However, while, from columns 2 and 3 of Table \ref{tab:SFR_table}, we do see that both the observed (SFR) and the theoretical maximum (\sfrmax) star formation rates increase when SN feedback is turned off (in both the HD and the MHD cases), the increase is, in both cases, only by a factors of a few to several, nowhere near the factor of two orders of magnitude required to resolve the star formation conundrum. Moreover, the increase factor is similar in both the SFR and \sfrmax, and therefore their ratio, \sfrmax/SFR (fourth column in Table \ref{tab:SFR_table}), barely changes, increasing only by $\sim 10\%$ in the Hydro models and even {\it decreasing} by $\sim 20\%$ in the MHD case when SN feedback is turned off. It is noteworthy that the theoretical maximum SFR is \textit{always} higher, by a factor of a few, when the SN feedback is turned off.

To further appreciate this, the depletion times for the calculated SFR, \sfrmax\ and the dense gas (bottom plot) are shown in Figure \ref{fig:dep_time} and Table \ref{tab:dep_times}. We see much shorter depletion times associated with \sfrmax\ compared to the total gas in all four simulations. This is to be expected, since \sfrmax\ is much larger than the actual SFR. Nevertheless, the ratio of $\tau_{\rm total}$ to $\tau_{\rm SFR_{max}}$ (last column of Table \ref{tab:dep_times}) is not the same as the corresponding (inverse) ratio between the star formation rates (fifth column in Table \ref{tab:SFR_table}). This reflects a variation in the total gas masses as well (cf.\ eq.\ \ref{depletion_time}).

This suggests that SN feedback mostly affects the availability of dense gas for star formation, rather than the rate at which it collapses to form stars. So, whilst it appears as though SN feedback controls the SFR, it is more likely that it is controlling the formation of dense gas, or gas reservoir needed for star formation. If there is more dense gas, there is more star formation, but the {\it specific} SFR of this dense gas (i.e., its rate of collapse) is not significantly altered.

\begin{table*} 
\centering
\begin{tabular}{l c c c c c c }
\hline
Model & $\tau_{\rm total}$ & $\tau_{\rm dense}$ & $\tau_{\rm _{SFR_{\rm max\,total}}}$ & $\tau_{\rm _{SFR_{\rm max}}}$ & $\tau_{\rm total}$ / & $\tau_{\rm dense}$ /\\
 & (yr) & (yr) & (yr) & (yr) & $\tau_{\rm SFR_{max\,total}}$ & $\tau_{\rm SFR_{max\,dense}}$ \\
 \\
\hline
Hydro          & $18.1\times 10^{9}$ & $2.08\times 10^{8}$ & $2.98\times 10^{8}$ & $1.74\times 10^{6}$ & 60.9 & 119.1 \\
Hydro\_no\_SNe & $2.51\times 10^{9}$ & $1.44\times 10^{8}$ & $0.41\times 10^{8}$ & $2.02\times 10^{6}$ & 61.8 & 71.6 \\
MHD            & $19.0\times 10^{9}$ & $2.48\times 10^{8}$ & $2.72\times 10^{8}$ & $1.37\times 10^{6}$ & 69.7 & 181.1 \\
MHD\_no\_SNe   & $5.31\times 10^{9}$ & $1.10\times 10^{8}$ & $1.04\times 10^{8}$ & $1.11\times 10^{6}$ & 50.8 & 98.5 \\
\hline
\end{tabular}
\caption{The average depletion times for the total gas and dense gas in the disc and based on the SFR$_{\rm max\,total}$ and SFR$_{\rm max}$ and their ratios.}
\label{tab:dep_times}
\end{table*}

\subsection{Controlling the formation of dense gas}
\label{dense_gas}

In this section we address two distinct questions. First, why does the ratio $\mathrm{SFR}/\mathrm{SFR}_{\max}$ remain nearly unchanged between simulations with and without SN feedback? Secondly, why is this ratio low in absolute terms in all four simulations, including those without SN feedback and magnetic support?

These questions are related, but not identical. The first concerns the near-invariance of the ratio across models, whereas the second concerns its absolute normalisation. We therefore discuss them separately below.

\subsubsection{Why does $\mathrm{SFR}/\mathrm{SFR}_{\max}$ remain nearly constant across the SN and no-SN runs?}

At first sight, one might expect the absence of SNe to produce a significantly larger $\mathrm{SFR}/\mathrm{SFR}_{\max}$. If so, one could worry that the observed near-constancy of this ratio is a numerical artefact of the sink-formation prescription. However, the key point is that we are addressing the \emph{constancy} of the ratio, not its absolute value. The sink prescription may affect the normalisation (i.e. the precise value measured in a given run), but this effect should act similarly in both SN and no-SN simulations, because the same numerical treatment is used in all cases. It is therefore unlikely, by itself, to explain why $\mathrm{SFR}/\mathrm{SFR}_{\max}$ remains approximately unchanged across feedback conditions.

This interpretation is supported by Fig.~\ref{fig:den_pdf}, which shows the volume-weighted density PDFs for gas with $n \geq 100~\mathrm{cm}^{-3}$ in all four simulations. The PDFs are broadly consistent with the power-law behaviour expected for collapsing gas \citep[e.g.,] [] {Ballesteros2011, Kritsuk2011, Girichidis2014} and remain continuous across the density at which sink-formation checks begin. Although removing SN feedback shifts more mass towards higher densities, the overall PDF shape does not show an artificial break or pile-up at the sink threshold. This suggests that the dense-gas distribution is still governed primarily by collapse dynamics, rather than imposed by the sink prescription.

It is also consistent with Table~\ref{tab:SFR_table}. When SN feedback is removed, the dense-gas mass increases by factors of $\sim 5.4$ (Hydro) and $\sim 1.5$ (MHD), while the SFR increases by factors of $\sim 7.3$ and $\sim 3.3$, respectively. The corresponding increase in specific SFR, ${\rm SFR}/M_{\rm dense}$, is therefore comparatively modest ($\sim 1.38$ in Hydro and $\sim 2.1$ in MHD). This supports a picture in which SN feedback primarily regulates the amount of dense gas available, rather than strongly altering the collapse rate per unit dense gas mass.

The same conclusion is supported by Table~\ref{tab:den_mf_table}, which reports the time-averaged dense-gas mass fractions in the intervals $10^2 \le n < 10^3~\mathrm{cm}^{-3}$ and $n \ge 10^3~\mathrm{cm}^{-3}$. In the SN runs, most dense gas lies in the lower interval (Hydro: $\sim 61\%$; MHD: $\sim 67\%$), whereas in Hydro\_no\_SNe the balance shifts towards higher densities ($\sim 57\%$ above $10^3~\mathrm{cm}^{-3}$), with MHD\_no\_SNe remaining more evenly split. This is a moderate redistribution, not a discontinuous pile-up at the sink threshold.

Moreover, if sink formation introduced a strong spurious effect, the most likely bias would be to remove gas from the dense phase too early, thereby {\it reducing} the measured dense-gas mass. This would act against, rather than induce, the enhanced dense-gas content seen in the no-SN runs. Hence, while the sink prescription may affect the absolute normalisation of $\mathrm{SFR}/\mathrm{SFR}_{\max}$, it is unlikely to explain its near-invariance between SN and no-SN cases.

\begin{figure}
	\includegraphics[width=\columnwidth]{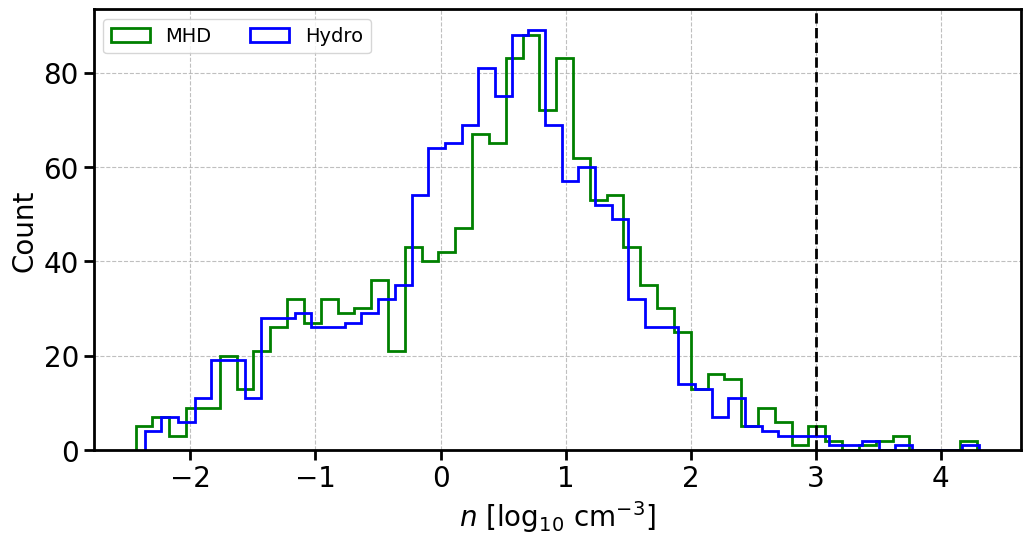}
    \caption{The mass-weighted mean density where SNe occur for the MHD (green) and Hydro (blue) models. We can see that almost all SNe occur in low density gas, $n\leq 100\,\rm cm^{-3}$.}
    \label{fig:sne}
\end{figure}

In addition, Fig.~\ref{fig:sne} shows that SNe occur predominantly in low-density gas in both the Hydro and MHD runs. This supports a picture in which SN feedback regulates star formation {\it indirectly}, by acting on the diffuse medium that supplies dense structures, rather than by directly suppressing collapse in the densest gas itself. In this context, the near-constant $\mathrm{SFR}/\mathrm{SFR}_{\max}$ across SN and no-SN cases is more naturally interpreted as a property of a similarly regulated dense-gas reservoir than as a sink-induced artefact.

Finally, whilst we do not fully exclude a contribution from rotation or shear to the global regulation of star formation, we note that rotation primarily acts on large scales, whereas the dense gas occupies much smaller regions. It is therefore unlikely to prevent the collapse of the densest gas directly. 

\subsubsection{Why is $\epsilon_{\rm ff}$ low in absolute terms?}

The more difficult question is why $\mathrm{SFR}/\mathrm{SFR}_{\max}$, or, equivalently, $\epsilon_{\rm ff}$, remains well below unity in all simulations, including those without SN feedback and magnetic support. The discussion above indicates that removing SNe changes the amount of dense gas far more than it changes $\epsilon_{\rm ff}$ itself, so the key issue is not the near-constancy of the ratio, but its low absolute value.

A plausible interpretation is that star formation is limited by a bottleneck in inward mass transport, rather than by direct support of the densest gas. Following \citet{Vazquez2024}, for structures with a spherically-averaged density profile ${\langle \rho(r) \rangle} \propto r^{-p}$ and $p<2$, the inward mass-transfer rate decreases towards smaller radii. In this regime, gas can continue to accumulate in dense structures, while only a reduced fraction is delivered to the stellar component per local free-fall time. In other words, collapse can proceed and the dense gas mass can grow, yet the final transfer from gas to stars remains throttled.

This can be expressed explicitly. If $M(r)\propto r^{3-p}$ and $\tau_{\rm ff}(\langle\rho{(r)}\rangle)\propto r^{p/2}$, then
\begin{equation}
    \mathrm{SFR}_{\max}(r)\equiv\frac{M(r)}{\tau_{\rm ff}(\langle\rho\rangle,r)}
    \propto r^{3(1-p/2)}.
    \label{eq:SFRmax_of_r}
\end{equation}
Thus, for $p<2$, $\mathrm{SFR}_{\max}(r)$ also declines inwards, mirroring the inward decline of the mass-transfer rate. This is consistent with a ``gravitational choking'' picture in which gas accumulates in the dense phase without being fully transferred to stars at the nominal maximum rate for the largest scales. We stress that this interpretation is physically motivated but still speculative in the context of the present simulations, which do not resolve the internal structure of the densest star forming gas or the small-scale inflow rates; testing it would require higher resolution and ideally Lagrangian tracers to measure the inward mass-transfer rate directly.

At the same time, we should not rule out other contributions to the absolute normalisation of $\epsilon_{\rm ff}$. As discussed above, the sink prescription may shift the normalisation of $\mathrm{SFR}/\mathrm{SFR}_{\max}$, even if it is unlikely to explain its near-invariance between SN and no-SN runs. In addition, geometrical effects may be important: collapse through filamentary networks can yield lower effective conversion rates than more isotropic collapse.

Finally, rotation and shear need not directly support the densest gas to influence $\epsilon_{\rm ff}$. At the same time, we should not rule out other contributions to the absolute normalisation of $\epsilon_{\rm ff}$. As discussed above, the sink prescription may shift the normalisation of $\mathrm{SFR}/\mathrm{SFR}_{\max}$, even if it is unlikely to explain its near-invariance between SN and no-SN runs. In addition, geometric effects may contribute to the low efficiency, since collapse through filamentary structures can lead to a less direct conversion of gas into stars than a more isotropic collapse \citep{Andre2017}.

\subsection{The role of the magnetic field on the SFR}

It has been suggested that magnetic fields reduce star formation in the ISM \citep{Krumholz2019,Hennebelle2019}, though to what extent and in what environments is still debated \citetalias{Whitworth2023}. Here, we find that when feedback is turned off, the global averaged field strength grows to a higher strength than with SNe, over a few tens of Myr. This, however, is not enough to counteract the increase in star formation generated by the increase in dense gas (see Section \ref{dense_gas} for more detail). The gravitational collapse is likely more powerful than the minimal additional pressure support being supplied. We do note that, compared to the Hydro simulation without feedback, the SFR is lower, by roughly half. So, whilst in a more realistic system with feedback there is little to no change in the SFR, we do see a reduction when feedback is removed.

Looking at the $B$-$n$ relationship, Fig.\ \ref{fig:B-n} compared to \citet{Crutcher2010} and \citet{whitworth2025}, neither simulation aligns. Notably, the loss of the diffuse, low field strength tail in the no feedback simulation highlights the loss of hot, diffuse gas driven by SNe. This is significant because it has been suggested that the hot, diffuse, turbulent gas controls the growth of the magnetic field by the small scale dynamo \citep{Gent2021,whitworth2025}. The lack of these low field strengths could be a result of insufficient resolution in the diffuse gas and thus not properly resolving this regime, (see Section 5.3.4 in \citet{whitworth2025} for a detailed discussion on this).

Figure \ref{fig:SFR} shows that, after a brief transient after turning the magnetic field off ($930 < t < 980$ Myr), the Hydro simulation (solid blue line) adjusts its SFR to essentially the same value as that of the MHD run (solid green line), and remains there for the rest of the integration time. Similarly, in Figure \ref{fig:dep_time} we see that the depletion time for the dense gas is virtually indistinguishable between the MHD and Hydro simulations. This suggests that the dense gas mass is also roughly the same in both cases, as indeed it can be seen in the left panel of Figure \ref{fig:SDRP}, which shows that both the Hydro and MHD runs including feedback have essentially the same dense gas radial profile.

These results suggest that the magnetic field has a negligible role in regulating the SFR in the presence of SN feedback. And, since we have already established in Sec.\ \ref{sec:SFR_tdepl} that the regulation of the SFR by the SNe is accomplished by regulating the dense gas mass, our results suggest then that, in the presence of SN driving, the magnetic field also has a negligible effect in regulating the dense gas mass. 

On the other hand, in the case with no SNe, we do see a larger SFR (by a factor $\sim 2$) for the Hydro simulation (cyan solid line in Figure \ref{fig:SFR}) than that for the corresponding MHD run (green solid line). However, the depletion time for the dense gas does not exhibit a corresponding variation from the Hydro to the MHD case (cyan dashed line and green dashed line, respectively in Figure \ref{fig:dep_time} and in Table \ref{tab:dep_times}), and instead the two depletion times are within $\sim 50\%$ from each other. This is because the Hydro simulation also has a much larger dense gas mass, which nearly compensates for the increased SFR, seen in the right panel of Figure \ref{fig:SDRP}.

These results suggest that, in the absence of feedback, the magnetic field does influence the SFR, by preventing the {\it formation} of dense gas by thermal condensation, presumably by increasing the total pressure of the diffuse gas, rather than by significantly modifying the collapse rate of the dense gas. Indeed, the SFR$_{\rm specific}$ of the dense gas (or, equivalently, its gas depletion time) remains essentially the same with and without the magnetic field, while the its mass is nearly 3 times larger in the Hydro case than in the MHD one (always without SNe), as also seen in the 7th column of Table \ref{tab:SFR_table}.

\section{Conclusions}
\label{conclusion}

In this paper we have presented four dwarf galaxy simulations with and without magnetic fields and with and without supernova feedback to determine what affects the galactic star formation rates and gas support more. We present our main conclusions here:

\begin{itemize}
    \item Although the total SFR does increase when SN feedback is turned off, so does the theoretical maximum \sfrmax, because the dense gas mass also increases. In all cases, \sfrmax\ remains nearly 1.75 orders of magnitude larger than the actual measured SFR. This implies that the discrepancy between the actual SFR and \sfrmax\ is {\it not due to SN feedback}.
    
    \item When SN feedback is turned off, the SFR per unit mass (SFR$_{\rm specific}$) increases only by a factor $\sim 1.38$ in the Hydro case and a factor $\sim 2.1$ in the MHD case.


    \item The depletion time of the dense gas remains within a factor of a few from that of \sfrmax, regardless of whether SNe are on or off, indicating that the efficiency of star formation per free‑fall time in the dense gas is nearly insensitive to the presence or absence of SN feedback.
    
    \item The above results indicate that the regulation of the SFR by the SN feedback is accomplished mainly by {\it regulating the conversion of diffuse to dense gas}, rather than by retarding the collapse of the dense gas.
    
    \item The effect of the magnetic field is negligible when SN feedback is on. However, when SNe are off, the presence of the magnetic field acts in a similar way to the feedback, by reducing the total amount of dense gas, suggesting that its main effect is to prevent the conversion of diffuse to dense gas, presumably because the magnetic pressure reduces the thermal condensation of the diffuse gas.
    
    \item The discrepancy between the observed and theoretical maximum SFRs persists in our simulations even when both SN feedback and magnetic fields are absent.

    \item The physical origin of the low $\epsilon_{\rm ff}$ in the dense gas remains unresolved here; gravitational choking is a plausible explanation, but requires highly resolved dense-gas structure with Lagrangian tracers to measure the inward mass-transfer rate.
    
\end{itemize}

\section*{Acknowledgements}

The authors thankfully acknowledge useful comments by an anonymous referee, in particular for pointing out potential problems that could arise from the sink-formation prescription, and the range of scales where rotation is most active. D.J.W.\ acknowledges support from the Programa de Becas Posdoctorales of the Direcci\'on General de Asuntos del Personal Acad\'{e}mico of the Universidad Nacional Aut\'{o}noma de M\'{e}xico (DGAPA,UNAM,Mexico) and support from the ANR CASCADE grant (ANR-24-ERCS-0004). E.V.-S.\ acknowledges financial support from UNAM-PAPIIT grant IG100223. J.B-P. acknowledges financial support from UNAM-PAPIIT grant IN110026. G.C.G.\ acknowledges support from UNAM-PAPIIT grant IN110824. 
This work used the DiRAC COSMA Durham facility managed by the Institute for Computational Cosmology on behalf of the STFC DiRAC HPC Facility (www.dirac.ac.uk). The equipment was funded by BEIS capital funding via STFC capital grants ST/P002293/1, ST/R002371/1 and ST/S002502/1, Durham University and STFC operations grant ST/R000832/1. DiRAC is part of the National e-Infrastructure.
This work made use of dedicated computing resources developed and maintained by the IRyA's Computing Department. The authors gratefully acknowledge the contributions of Daniel Díaz-González, Miguel Espejel, Alfonso Ginori González, and Gilberto Zavala.

\section*{Data Availability}

All data is available upon request to the first author.



\bibliographystyle{mnras}
\bibliography{ref} 




\appendix




\bsp	
\label{lastpage}
\end{document}